\documentstyle[12pt,emlines]{article}
\newcommand{\dfrac}{\displaystyle \frac}
\newcommand{\Dsla}{D\hspace{-9.25pt} /}
\def\x{\hspace{1ex}}
\def\ao{{}\kern-.10em\hbox{``}}
\begin{document}
\large
\bibliographystyle{plain}
\begin{titlepage}
\hfill \begin{tabular}{l}HEPHY-PUB 581/93\\ UWThPh-1993-34\\ August 1993
\end{tabular}\\[4cm]
\begin{center}
{\Large \bf QUADRATIC DIVERGENCES IN GUTS WITH VANISHING ONE-LOOP BETA
FUNCTIONS}\\
\vspace{1.5cm}
{\Large \bf Wolfgang LUCHA}\\[.5cm]
{\large Institut f\"ur Hochenergiephysik\\
\"Osterreichische Akademie der Wissenschaften\\
Nikolsdorfergasse 18, A-1050 Wien, Austria}\\[1cm]
{\Large \bf Michael MOSER}\\[.5cm]
{\large Institut f\"ur Theoretische Physik\\
Universit\"at Wien\\
Boltzmanngasse 5, A-1090 Wien, Austria}\\[3cm]
{\bf Abstract}
\end{center}
\normalsize

All members of a recently proposed new set of (non-supersymmetric) grand
unified theories with at the one-loop level vanishing beta functions for the
gauge, Yukawa, and scalar-boson self-interaction coupling constants are shown
to involve, already at the one-loop level, quadratically divergent
contributions to both the vector-boson and scalar-boson masses.
\end{titlepage}

\section{Introduction}

Supersymmetry rendered possible the construction of finite quantum field
theories, i.~e., quantum field theories which in their perturbation expansion
are free from ultraviolet divergences, at least up to two loops \cite{N=1}
for (even softly broken \cite{N=1softbreak}) $N = 1$ supersymmetric theories
satisfying two so-called \ao finiteness conditions" and at all orders
\cite{howe83a} for (again even softly broken \cite{N=2softbreak}) $N = 2$
supersymmetric theories satisfying a single one-loop finiteness condition.
The latter thus form a large class of finite quantum field theories
\cite{N=2class} which includes the famous $N = 4$ super-Yang--Mills theory as
a special case \cite{N=4SYM}.

The discovery of these supersymmetric finite quantum field theories suggested
to ask oneself whether supersymmetry is indeed a necessary prerequisite for
finiteness in the sense described above \cite{lucha86a,lucha86b,lucha87a}
and, in particular, prompted the search for non-supersymmetric finite models.
However, the conclusions at which one arrives unfortunately depend on the
chosen regularization method: Dimensional regularization, for a fixed
space--time dimension, ignores all quadratic divergences in the theory. In
contrast to this, a regularization method which employs a dimensional
regularization parameter like, e.~g., cutoff regularization, allows to
identify also this latter type of divergence, being thus by far more
restrictive when demanding absence of divergences. Accordingly, all
non-supersymmetric but \ao dimensional-regularization finite" models given so
far \cite{boehm87} proved to entail quadratic divergences \cite{lucha87b}.

Very recently, some new set of models has been singled out by the requirement
of vanishing one-loop $\beta$ functions for all (gauge, Yukawa, and
scalar-boson self-interaction) coupling constants in some massless theory
with specific non-supersymmetric particle content \cite{shapiro93}.

In the present note we would like to demonstrate that the models proposed in
Ref. \cite{shapiro93}, in spite of the fact that they do not possess any
one-loop divergences when investigated by dimensional regularization, produce
quadratically divergent one-loop contributions to the masses of both the
vector bosons and the scalar bosons present in the theory. To this end we
discuss, in Sect. \ref{sec:qdgength}, the quadratic divergences arising in a
general gauge theory at the one-loop level. In Sect. \ref{sec:psfinmod} we
review the one-loop finite models of Ref. \cite{shapiro93} and, by
application of the previous general discussion, analyse these models with
respect to their eventual quadratic divergences. We are forced to conclude,
in Sect. \ref{sec:shayagco}, that all of these models, already at one-loop
level, involve quadratic divergences.

\newpage

\section{Quadratic Divergences in a General Gauge Theory}\label{sec:qdgength}

A gauge theory is characterized by invariance with respect to local
transformations forming some compact---in general, non-Abelian---gauge group
${\cal G}$, which is defined in terms of its (completely antisymmetric)
structure constants $f_{abc}$ by the commutation relations $[T^a,T^b] = i
f_{abc} T^c$ of its (Hermitean) generators $T^a$. Upon ignoring all
dimensional parameters like masses and cubic self-couplings of scalar bosons
(which do not affect the high-energy behaviour of the theory), the Lagrangian
defining a general gauge theory for
\begin{itemize}
\item (Hermitean) vector gauge fields $V_\mu^a$ in the adjoint representation
$G$ of ${\cal G}$,
\item two-component Weyl spinor fields $\psi_{iL} \equiv \frac{1}{2}\left(1 +
\gamma_5\right)\psi_i$ in some fermion representation $F$ of ${\cal G}$
(because of $\left(\psi_R\right)^c = \psi^c{}_L$ without loss of generality
all of them assumed to be, say, left-chiral), and
\item Hermitean scalar fields $\phi_m$ in some (necessarily) real
scalar-boson representation $S$ of ${\cal G}$
\end{itemize}
is given by
\begin{eqnarray}
{\cal L}
&=& - \frac{1}{4} F_{\mu\nu}^a F^{\mu\nu}_a
+ i \overline{\psi_L} \gamma^\mu D_\mu \psi_L
+ \frac{1}{2} \left(D_\mu\phi\right)^T D^\mu\phi \nonumber\\
&+& \frac{1}{2} \left[\overline{\psi^c{}_R} h^m \psi_L \phi_m
+ \mbox{H. c.}\right]
- \frac{1}{4!} \lambda_{mnpq} \phi_m \phi_n \phi_p \phi_q \quad ,
\end{eqnarray}
where $F_{\mu\nu}^a$ denotes the gauge-covariant field strength
\begin{equation}
F_{\mu\nu}^a \equiv \partial_\mu V_\nu^a - \partial_\nu V_\mu^a
+ g f_{abc} V_\mu^b V_\nu^c
\end{equation}
and $D_\mu$ the gauge-covariant derivative
\begin{equation}
D_\mu \equiv \partial_\mu - i g V_\mu^a T^a_R \quad , \qquad R = F,S \quad ,
\end{equation}
both of them introducing gauge interactions with coupling strength $g$.
Permutation symmetry and gauge invariance impose some obvious restrictions on
the Yukawa coupling matrices $h^m_{ik}$ as well as on the quartic
self-couplings $\lambda_{mnpq}$ of the scalar bosons.

On dimensional grounds, for the above theory quadratic divergences can only
arise in the masses of vector bosons---which would destroy gauge
invariance---and in the masses of scalar bosons. Apart from some trivial
factors, these quadratically divergent contributions to vector and
scalar-boson masses are proportional to some quantities $Q_V$ and $Q_S$,
respectively, which at the one-loop level, when expressed in terms of the
quadratic Casimir operator
\begin{equation}
\sum_\sigma C_2(R_\sigma) E^\sigma_{ik} := \left(T^a_R T^a_R\right)_{ik}
\end{equation}
(where $E^\sigma$ denotes the projector onto the irreducible component
$R_\sigma$ in the decomposition $R = \bigoplus_\sigma R_\sigma$ of the
maybe reducible representation $R$) and the second-order Dynkin index
\begin{equation}
T(R)\delta_{ab} := \mbox{Tr}(T^a_R T^b_R) \quad , \qquad
T(R) = \sum_\sigma T(R_\sigma) \quad ,
\end{equation}
read for the mass of the vector bosons \cite{lucha86a}
\begin{equation}
Q_V = 2 \, C_2(G) - 2 \, T(F) + T(S)
\label{eq:Q_V}
\end{equation}
and for the mass of the scalar bosons \cite{lucha86a,lucha86b,lucha87b}
\begin{equation}
\left(Q_S\right)_{mn} = 6 \, g^2 \left(T^a_S T^a_S\right)_{mn}
- 4 \, \mbox{Re\ Tr} \left(h^m{h^n}^\dagger\right) + \lambda_{mnpp} \quad .
\label{eq:Q_S}
\end{equation}
In the next section we show that all models given in Ref. \cite{shapiro93},
although possessing vanishing one-loop $\beta$ functions, yield non-vanishing
values for the quantities $Q_V$ and $Q_S$, and thus have to be regarded as
merely \ao pseudo-finite".

\section{The Pseudo-Finite Models}\label{sec:psfinmod}

Both of the models under consideration are based on the gauge group ${\cal G}
= \mbox{SU($N$)}$ and involve only particles transforming either according to
the fundamental representation $R_f$ (of dimension $d_f = N$) or according to
the adjoint representation $G$ (of dimension $d_G = N^2 - 1$) of ${\cal G}$.
In order to be able to fulfill the requirements of the assumed gauge
invariance, both models have to contain (real) gauge vector bosons $V_\mu^a$,
$a = 1,2,\dots , N^2 - 1$, transforming, of course, according to the adjoint
representation $G$ of the gauge group ${\cal G}$, i.~e.,
\begin{equation}
V_\mu \sim G \quad .
\end{equation}
As a consequence, in both models all couplings may be expressed in terms of
the generators $(T^a_G)_{bc} = \frac{1}{i} f_{abc}$ in the adjoint
representation $G$ of ${\cal G}$, the generators $T_f$ in the fundamental
representation $R_f$ of ${\cal G}$, or the completely symmetric constants
$d_{abc} \equiv \mbox{Tr} \left(\left\{T_f^a,T_f^b\right\}T_f^c\right) / \:
T(R_f)$.

\subsection{The general model}

The non-vector particle content of this model consists of
\begin{itemize}
\item $m$ sets of Dirac fermions $\Psi_{(k)}$, $k = 1,2,\dots ,m$, each of
these sets transforming according to the adjoint representation $G$ of
${\cal G}$, i.~e.,
\begin{equation}
\Psi_{(k)} \sim G \quad , \qquad k = 1,2,\dots ,m \quad ;
\label{eq:shayag-Psi}
\end{equation}
\item $m$ sets of Dirac fermions $\chi_{(k)}$, $k = 1,2,\dots ,m$, each of
these sets transforming according to the fundamental representation $R_f$ of
${\cal G}$, i.~e.,
\begin{equation}
\chi_{(k)} \sim R_f \quad , \qquad k = 1,2,\dots ,m \quad ;
\end{equation}
\item $n$ sets of Dirac fermions $\zeta_{(k)}$, $k = 1,2,\dots ,n$, each of
these sets transforming according to the fundamental representation $R_f$ of
${\cal G}$, i.~e.,
\begin{equation}
\zeta_{(k)} \sim R_f \quad , \qquad k = 1,2,\dots ,n \quad ;
\label{eq:shayag-chi}
\end{equation}
\item real scalar bosons $\Phi^a$, $a = 1,2,\dots , N^2 - 1$, transforming
according to the adjoint representation $G$ of ${\cal G}$, i.~e.,
\begin{equation}
\Phi \sim G \quad ;
\end{equation}
\item (necessarily) complex scalar bosons $\varphi$ transforming
according to the fundamental representation $R_f$ of ${\cal G}$, i.~e.,
\begin{equation}
\varphi \sim R_f \quad .
\label{eq:shayag-varphi}
\end{equation}
\end{itemize}
The Lagrangian defining this general model reads \cite{shapiro93}
\begin{eqnarray}
{\cal L}
&=& - \frac{1}{4} F_{\mu\nu}^a F^{\mu\nu}_a
+ i \sum_{k = 1}^m \bar\Psi_{(k)}^a \left(\Dsla_{ab} - h_1 f_{abc}
\Phi^c\right) \Psi_{(k)}^b \nonumber\\
&+& i \sum_{k = 1}^m \bar\chi_{(k)} \left(\Dsla - i h_2 T_f^a\Phi^a\right)
\chi_{(k)}
+ i \sum_{k = 1}^n \bar\zeta_{(k)} \Dsla \: \zeta_{(k)} \nonumber\\
&+& \left(i h_3 \sum_{k = 1}^m \bar\chi_{(k)} T_f^a \varphi \Psi_{(k)}^a +
\mbox{H. c.}\right) \nonumber\\
&+& \frac{1}{2} \left(D_\mu \Phi\right)^T D^\mu \Phi
+ \left(D_\mu \varphi\right)^\dagger D^\mu \varphi
- \frac{\lambda_1}{8} \left(\Phi^T \Phi\right)^2
- \frac{\lambda_2}{8} \left(\Phi^a d_{abc} \Phi^b\right)^2 \nonumber\\
&-& \frac{\lambda_3}{2} \left(\Phi^T \Phi\right)\left(\varphi^\dagger
\varphi\right)
- \frac{\lambda_4}{2} \left(\Phi^a d_{abc} \Phi^b\right)\left(\varphi^\dagger
T_f^c \varphi\right)
- \frac{\lambda_5}{2} \left(\varphi^\dagger \varphi\right)^2 \quad .
\label{eq:Lagr-genmod}
\end{eqnarray}
The fermions $\chi_{(k)}$ are discriminated from the fermions $\zeta_{(k)}$ by
the fact that the former also undergo Yukawa interactions whereas the latter
do not.

Finiteness of the one-loop contribution to the renormalization of the gauge
coupling constant, as expressed by the relation
\begin{equation}
21 N - 4 [(2 N + 1) m + n] = 1 \quad ,
\label{eq:olf-gc/gen}
\end{equation}
restricts the possible gauge groups SU($N$) to the values $N = 4 \ell + 1$
for $\ell = 1,2,\dots$. The multiplicities $m$ and $n$ allowed by Eq.
(\ref{eq:olf-gc/gen}) for the groups SU(5) and SU(9) are listed, together
with the respective number of solutions\footnote{\normalsize\ By inspection,
the $(N = 5, m = 2, n = 4)$ model of Ref. \cite{shapiro93}, although
attributed by the authors to be one-loop finite, turns out---at least for the
numerical values of the Yukawa interactions and quartic scalar-boson
self-couplings given in Ref. \cite{shapiro93}---to possess non-vanishing
one-loop $\beta$ functions and thus to be not even pseudo-finite.} of the
one-loop finiteness conditions for Yukawa interactions and quartic
scalar-boson self-couplings, in Table \ref{tab:mult-gen}.

\begin{table}[h]
\begin{center}
\caption{Multiplicities $m$ and $n$ allowed by one-loop finiteness of the
gauge coupling constant for general models based on the smallest conceivable
gauge groups SU(5) and SU(9), and corresponding number of solutions of the
one-loop finiteness conditions for the Yukawa interactions and quartic
scalar-boson self-couplings}\label{tab:mult-gen}
\vspace{0.5cm}
\begin{tabular}{|r|r|r|c|}
\hline
&&&\\[-1ex]
\multicolumn{1}{|c|}{$N$}&\multicolumn{1}{c|}{$m$}&\multicolumn{1}{c|}{$n$}&
\begin{tabular}{l}number of\\ solutions\end{tabular}\\
&&&\\[-1.5ex]
\hline\hline
&&&\\[-1.5ex]
\x\x\x 5\x\x\x&\x\x\x 0\x\x\x&\x\x\x 26\x\x\x&0\\
&\x\x\x 1\x\x\x&\x\x\x 15\x\x\x&1\\
&\x\x\x 2\x\x\x&\x\x\x 4\x\x\x&0\\
&&&\\[-1.5ex]
\hline
&&&\\[-1.5ex]
\x\x\x 9\x\x\x&\x\x\x 0\x\x\x&\x\x\x 47\x\x\x&0\\
&\x\x\x 1\x\x\x&\x\x\x 28\x\x\x&1\\
&\x\x\x 2\x\x\x&\x\x\x 9\x\x\x&1\\[1.5ex]
\hline
\end{tabular}
\end{center}
\end{table}

The numerical values of the Yukawa coupling constants $h_1,h_2,h_3$ and of
the scalar-boson self-coupling constants $\lambda_1,\lambda_2,\dots
,\lambda_5$ which render the three models filtered out by the analysis in
Ref. \cite{shapiro93} finite at the one-loop level are compiled in Table
\ref{tab:numval}.

\begin{table}[hbt]
\begin{center}
\caption[]{Numerical values \cite{shapiro93} of the Yukawa coupling constants
$h_i^2$, $i = 1,2,3$, and of the scalar-boson self-coupling constants
$\lambda_i$, $i = 1,2,\dots ,5$, for the general pseudo-finite models
identified in Table \ref{tab:mult-gen} as well as the resulting values of the
quantity $Q_S$, which parametrizes the magnitude of the quadratically
divergent one-loop contribution to the scalar-boson masses, for both the
sectors of the scalar bosons $\Phi$ and $\varphi$, denoted in these sectors
by $Q_S^{(\Phi)}$ and $Q_S^{(\varphi)}$, respectively. (All of these
quantities in units of $g^2$.)}\label{tab:numval}
\vspace{0.5cm}
\begin{tabular}{|c|c|c|c|}
\hline
&&&\\[-1ex]
$\quad$Model$\quad$&
\begin{tabular}{c}$N = 5$\\$m = 1,\ n = 15$\end{tabular}&
\begin{tabular}{c}$N = 9$\\$m = 1,\ n = 28$\end{tabular}&
\begin{tabular}{c}$N = 9$\\$m = 2,\ n = 9$\end{tabular}\\
&&&\\[-1.5ex]
\hline\hline
&&&\\[-1.5ex]
$h_1^2/g^2$&1.4211\dots&1.4546\dots&0.9817\dots\\
$h_2^2/g^2$&1.6806\dots&1.7416\dots&0.3878\dots\\
$h_3^2/g^2$&2.3612\dots&2.3294\dots&1.1273\dots\\
&&&\\[-1.5ex]
\hline
&&&\\[-1.5ex]
$\lambda_1/g^2$&0.6594\dots&0.4149\dots&0.3685\dots\\
$\lambda_2/g^2$&1.2933\dots&1.1947\dots&0.6880\dots\\
$\lambda_3/g^2$&0.3235\dots&0.1756\dots&0.0899\dots\\
$\lambda_4/g^2$&1.6765\dots&1.7329\dots&0.9858\dots\\
$\lambda_5/g^2$&1.0385\dots&1.1369\dots&0.6088\dots\\
&&&\\[-1.5ex]
\hline\hline
&&&\\[-1.5ex]
$Q_S^{(\Phi)}/g^2$&\x$- 2.32$&\x$- 0.07$&$- 46.86$\\
$Q_S^{(\varphi)}/g^2$&$- 10.71$&$- 19.37$&$- 34.13$\\[1.5ex]
\hline
\end{tabular}
\end{center}
\end{table}

It is, however, an easy task to convince oneself that for the present general
model the quantities $Q_V$ and $Q_S$ as defined in Eqs. (\ref{eq:Q_V}) and
(\ref{eq:Q_S}), which parametrize the magnitude of the one-loop contribution
to the quadratic divergence of the vector-boson and scalar-boson masses,
respectively, are definitely non-vanishing:
\begin{itemize}
\item $Q_V$ is given by
\begin{equation}
Q_V = 3 N - 2 [(2 N + 1) m + n] + 1
\label{eq:QV-genmod}
\end{equation}
or---after elimination of the fermion contribution with the help of the
one-loop finiteness condition (\ref{eq:olf-gc/gen}) for the gauge coupling
constant---by
\begin{equation}
Q_V = - \frac{15 N -3}{2} \quad ,
\end{equation}
which is beyond doubt unequal to zero for any integer $N$ and, in fact,
strictly negative for all $N = 1,2,\dots$.
\item $Q_S$ is given in the sector of the scalar bosons $\Phi$ by
\begin{eqnarray}
Q_S^{(\Phi)}
&=& 6 N g^2 - 4 m \left(2 N h_1^2 + h_2^2\right) \nonumber\\
&+& \left(N^2 + 1\right) \lambda_1 + 2 \, \dfrac{N^2 - 4}{N} \, \lambda_2
+ 2 N \lambda_3
\label{eq:QS-Phi}
\end{eqnarray}
and in the sector of the scalar bosons $\varphi$ by
\begin{eqnarray}
Q_S^{(\varphi)}
&=& 3 \, \dfrac{N^2 - 1}{N} \, g^2 - 4 m \, \dfrac{N^2 - 1}{N} \, h_3^2
\nonumber\\
&+& \left(N^2 - 1\right) \lambda_3 + 2 \left(N + 1\right) \lambda_5 \quad .
\label{eq:QS-varphi}
\end{eqnarray}
Evaluation of the right-hand sides of Eqs. (\ref{eq:QS-Phi}) and
(\ref{eq:QS-varphi}) with the help of the three sets of solutions for the
Yukawa interactions and scalar-boson self-couplings quoted in Table
\ref{tab:numval} yields the numerical results for $Q_S^{(\Phi)}$ and
$Q_S^{(\varphi)}$ given also in Table \ref{tab:numval}. The non-vanishing
values of these quantities indicate unambiguously that in each of the general
one-loop pseudo-finite models of Table \ref{tab:mult-gen} there arise
quadratic divergences for the masses of the scalar bosons.
\end{itemize}

\subsection{The simplified model}

This model is obtained from the more general model described above by
completely decoupling the fermions $\Psi_{(k)}$, Eq. (\ref{eq:shayag-Psi}),
and $\zeta_{(k)}$, Eq. (\ref{eq:shayag-chi}), as well as the scalar bosons
$\varphi$, Eq. (\ref{eq:shayag-varphi}), from the theory. Accordingly, the
non-vector particle content of this model consists of
\begin{itemize}
\item $m$ sets of Dirac fermions $\chi_{(k)}$, $k = 1,2,\dots ,m$, each of
these sets transforming according to the fundamental representation $R_f$ of
${\cal G}$, i.~e.,
\begin{equation}
\chi_{(k)} \sim R_f \quad , \qquad k = 1,2,\dots ,m \quad ;
\end{equation}
\item real scalar bosons $\Phi^a$, $a = 1,2,\dots , N^2 - 1$, transforming
according to the adjoint representation $G$ of ${\cal G}$, i.~e.,
\begin{equation}
\Phi \sim G \quad .
\end{equation}
\end{itemize}
Consequently, the Lagrangian defining this simplified model reads
\begin{eqnarray}
{\cal L}
&=& - \frac{1}{4} F_{\mu\nu}^a F^{\mu\nu}_a
+ i \sum_{k = 1}^m \bar\chi_{(k)} \left(\Dsla - i h T_f^a\Phi^a\right)
\chi_{(k)} \nonumber\\
&+& \frac{1}{2} \left(D_\mu \Phi\right)^T D^\mu \Phi
- \frac{\lambda_1}{8} \left(\Phi^T \Phi\right)^2
- \frac{\lambda_2}{8} \left(\Phi^a d_{abc} \Phi^b\right)^2 \quad .
\end{eqnarray}

Finiteness of the one-loop contribution to the renormalization of the gauge
coupling constant, as expressed in this case by the relation
\begin{equation}
21 N = 4 m \quad ,
\label{eq:olf-gc/simpl}
\end{equation}
now restricts the possible gauge groups SU($N$) to the values $N = 4 \ell$
for $\ell = 1,2,\dots$. The multiplicity $m$ fixed by Eq.
(\ref{eq:olf-gc/simpl}) for the groups SU(4) and SU(8) is listed, together
with the respective number of solutions of the one-loop finiteness conditions
for Yukawa interactions and quartic scalar-boson self-couplings, in Table
\ref{tab:mult-simpl}. According to this, there exist two solutions for the
$(N = 8, m = 42)$ model whereas there are none for the $(N = 4, m = 21)$
model. Both of these solutions are characterized by vanishing Yukawa
interactions, i.~e., by $h = 0$ \cite{shapiro93}.

\begin{table}[hbt]
\begin{center}
\caption{Multiplicity $m$ as fixed by one-loop finiteness of the gauge
coupling constant for simplified models based on the smallest conceivable
gauge groups SU(4) and SU(8), and corresponding number of solutions of the
one-loop finiteness conditions for the Yukawa interactions and quartic
scalar-boson self-couplings}\label{tab:mult-simpl}
\vspace{0.5cm}
\begin{tabular}{|r|r|c|}
\hline
&&\\[-1ex]
\multicolumn{1}{|c|}{$N$}&\multicolumn{1}{c|}{$m$}&
\begin{tabular}{l}number of\\ solutions\end{tabular}\\
&&\\[-1.5ex]
\hline\hline
&&\\[-1.5ex]
\x\x\x 4\x\x\x&\x\x\x 21\x\x\x&0\\
&&\\[-1.5ex]
\hline
&&\\[-1.5ex]
\x\x\x 8\x\x\x&\x\x\x 42\x\x\x&2\\[1.5ex]
\hline
\end{tabular}
\end{center}
\end{table}

Again it is straightforward to check whether or not the quantities $Q_V$ and
$Q_S$, which characterize the one-loop quadratic divergences in vector- and
scalar-boson masses, respectively, vanish:
\begin{itemize}
\item $Q_V$ reduces from Eq. (\ref{eq:QV-genmod}), valid for the general
model (\ref{eq:Lagr-genmod}), to
\begin{equation}
Q_V = 3 N - 2 m
\end{equation}
or---when replacing the multiplicity $m$ by the expression resulting from the
one-loop finiteness condition (\ref{eq:olf-gc/simpl}) for the gauge coupling
constant---to
\begin{equation}
Q_V = - \frac{15 N}{2} \quad ,
\end{equation}
which obviously is strictly negative for all, in any case positive, $N$.
\item Without any further calculation, $Q_S$ may be read off immediately from
the corresponding expression $Q_S^{(\Phi)}$ for the scalar bosons $\Phi$ of
the general model, given in Eq. (\ref{eq:QS-Phi}), by just dropping those
contributions which arise, on the one hand, from the Yukawa coupling
proportional to $h_1$ and, on the other hand, from the scalar-boson
self-interaction proportional to $\lambda_3$ (and by re-labelling $h_2$
simply by $h$):
\begin{equation}
\left(Q_S\right)_{ab} =: Q_S \delta_{ab}
\end{equation}
with
\begin{equation}
Q_S = 6 N g^2 - 4 m h^2
+ \left(N^2 + 1\right) \lambda_1 + 2 \, \frac{N^2 - 4}{N} \, \lambda_2 \quad .
\end{equation}
Since, for reasons of stability of the theory, all $\lambda_i$, $i = 1,2$,
have to be positive, this last expression is for $h = 0$, irrespective of the
precise numerical values of the couplings $\lambda_i$, strictly positive.
\end{itemize}

\section{Conclusion}\label{sec:shayagco}

Two recently proposed sets of grand unified theories, characterized by a
definite choice of some specific non-supersymmetric particle content and the
fact that the beta functions of their gauge, Yukawa, and scalar-boson
self-interaction coupling constants vanish at the one-loop level, have been
investigated with respect to the eventual appearance of quadratic divergences
in the course of renormalization of vector-boson and scalar-boson masses,
respectively. Both of these two sets of models are based on the gauge group
SU($N$); the more general one involves fermions and scalar bosons in (some
multiples of) the fundamental and the adjoint representation of SU($N$), the
rather simplified one contains only fermions in some multiple of the
fundamental representation and scalar bosons in the adjoint representation of
SU($N$). The requirement of vanishing one-loop beta functions fixes the
possible gauge groups, i.~e., $N$, the multiplicities of all the fermion
representations, as well as the numerical values of the Yukawa and
scalar-boson self-interaction coupling constants. The resulting models thus
appear to be one-loop finite when all divergences are handled by dimensional
regularization.

This situation, however, may change completely when employing a
regularization method which operates with a dimensional regulator. And
indeed, in the present analysis we were able to show that in each of the
above models both the vector-boson and scalar-boson masses receive
quadratically divergent contributions at the one-loop level. In other words,
all of these models are plagued by quadratic divergences and, consequently,
should not be regarded to be one-loop finite in a regularization-scheme
independent manner.

Moreover, by considering the condition for two-loop finiteness of the gauge
coupling constant \cite{lucha86a,lucha86b,boehm87}, all of the above models
can easily be shown to loose their pseudo-finiteness at the two-loop level,
as has been suspected already by their authors themselves \cite{shapiro93}.


\newpage

\normalsize

\begin{thebibliography}{99}
\bibitem{N=1}
 {\sc D.~R.~T.~Jones} and {\sc L.~Mezincescu}, Phys. Lett. {\bf 136~B} (1984)
 242;\\
 {\sc P.~West}, Phys. Lett. {\bf 137~B} (1984) 371;\\
 {\sc A.~Parkes} and {\sc P.~West}, Phys. Lett. {\bf 138~B} (1984) 99;\\
 {\sc D.~R.~T.~Jones} and {\sc L.~Mezincescu}, Phys. Lett. {\bf 138~B} (1984)
 293.
\bibitem{N=1softbreak}
 {\sc D.~R.~T.~Jones}, {\sc L.~Mezincescu}, and {\sc Y.-P.~Yao}, Phys. Lett.
 {\bf 148~B} (1984) 317;\\
 {\sc J.~Le\'{o}n} and {\sc J.~P\'{e}rez-Mercader}, Phys. Lett. {\bf 164~B}
 (1985) 95.
\bibitem{howe83a}
 {\sc P.~S.~Howe}, {\sc K.~S.~Stelle}, and {\sc P.~C.~West}, Phys. Lett. {\bf
 124~B} (1983) 55.
\bibitem{N=2softbreak}
 {\sc A.~Parkes} and {\sc P.~West}, Phys. Lett. {\bf 127~B} (1983) 353;\\
 {\sc J.-M.~Fr\`{e}re}, {\sc L.~Mezincescu}, and {\sc Y.-P.~Yao}, Phys. Rev.
 D {\bf 29} (1984) 1196;\\
 {\sc J.-M.~Fr\`{e}re}, {\sc L.~Mezincescu}, and {\sc Y.-P.~Yao}, Phys. Rev.
 D {\bf 30} (1984) 2238.
\bibitem{N=2class}
 {\sc I.~G.~Koh} and {\sc S.~Rajpoot}, Phys. Lett. {\bf 135~B} (1984) 397;\\
 {\sc F.-x.~Dong}, {\sc T.-s.~Tu}, {\sc P.-y.~Xue}, and {\sc X.-j.~Zhou},
 Phys. Lett. {\bf 140~B} (1984) 333;\\
 {\sc J.-P.~Derendinger}, {\sc S.~Ferrara}, and {\sc A.~Masiero}, Phys. Lett.
 {\bf 143~B} (1984) 133;\\
 {\sc X.-d.~Jiang} and {\sc X.-j.~Zhou}, Phys. Lett. {\bf 144~B} (1984) 370;\\
 {\sc S.~Kalara}, {\sc D.~Chang}, {\sc R.~N.~Mohapatra}, and {\sc
 A.~Gangopadhyaya}, Phys. Lett. {\bf 145~B} (1984) 323;\\
 {\sc P.~Fayet}, Phys. Lett. {\bf 153~B} (1985) 397.
\bibitem{N=4SYM}
 {\sc P.~S.~Howe}, {\sc K.~S.~Stelle}, and {\sc P.~K.~Townsend}, Nucl. Phys.
 {\bf B~214} (1983) 519;\\
 {\sc P.~S.~Howe}, {\sc K.~S.~Stelle}, and {\sc P.~K.~Townsend}, Nucl. Phys.
 {\bf B~236} (1984) 125.
\bibitem{lucha86a}
 {\sc W.~Lucha} and {\sc H.~Neufeld}, Phys. Rev. D {\bf 34} (1986) 1089.
\bibitem{lucha86b}
 {\sc W.~Lucha} and {\sc H.~Neufeld}, Phys. Lett. B {\bf 174} (1986) 186.
\bibitem{lucha87a}
 {\sc W.~Lucha} and {\sc H.~Neufeld}, Helvetica Physica Acta {\bf 60} (1987)
 699.
\bibitem{boehm87}
 {\sc M.~B\"ohm} and {\sc A.~Denner}, Nucl. Phys. {\bf B~282} (1987) 206.
\bibitem{lucha87b}
 {\sc W.~Lucha}, Phys. Lett. B {\bf 191} (1987) 404.
\bibitem{shapiro93}
 {\sc I.~L.~Shapiro} and {\sc E.~G.~Yagunov}, Int. J. Mod. Phys. A {\bf 8}
 (1993) 1787.
\end{thebibliography}
\end{document}